\documentclass[trackchanges]{aastex7}
\DeclareUnicodeCharacter{02BC}{\textquotesingle}

\usepackage{graphicx}
\usepackage{subcaption}
\usepackage{amsmath}
\usepackage{bm}

\begin{document}

\title{LSTM-MDNz: Estimating Quasar Photometric Redshifts with an LSTM-Augmented Mixture Density Network}

\author[orcid=0009-0005-1876-699X, gname=Jianzhen, sname=Chen]{Jianzhen Chen} 
\affiliation{Shanghai Key Lab for Astrophysics, Shanghai Normal University, Shanghai 200234, People’s Republic of China}
\email{jzchen@shnu.edu.cn}

\author[orcid=0009-0009-1617-8747,sname=Luo]{Zhijian Luo}
\affiliation{Shanghai Key Lab for Astrophysics, Shanghai Normal University, Shanghai 200234, People’s Republic of China}
\email[show]{zjluo@shnu.edu.cn} 

%\author{Zhirui Tang}
%\affiliation{Shanghai Key Lab for Astrophysics, Shanghai Normal University, Shanghai 200234, People’s Republic of China}
%\email{junhou@shnu.edu.cn}

\author[orcid=0000-0003-0688-8445]{Liping Fu}
\affiliation{Shanghai Key Lab for Astrophysics, Shanghai Normal University, Shanghai 200234, People’s Republic of China}
\email{fuliping@shnu.edu.cn}

\author[orcid=0000-0002-2326-0476]{Zhu Chen}
\affiliation{Shanghai Key Lab for Astrophysics, Shanghai Normal University, Shanghai 200234, People’s Republic of China}
\email{zhuchen@shnu.edu.cn}

\author[0000-0001-8244-1229,sname=Xiao,gname=Hubing]{Hubing Xiao}
\affiliation{Shanghai Key Lab for Astrophysics, Shanghai Normal University, Shanghai 200234, People’s Republic of China}
\email{hubing.xiao@outlook.com}

\author[0000-0002-2326-0476,sname=Zhang,gname=Shaohua]{Shaohua Zhang}
\affiliation{Shanghai Key Lab for Astrophysics, Shanghai Normal University, Shanghai 200234, People’s Republic of China}
\email{zhangshaohua@shnu.edu.cn}

\author[gname=Chenggang]{Chenggang Shu}
\affiliation{Shanghai Key Lab for Astrophysics, Shanghai Normal University, Shanghai 200234, People’s Republic of China}
\email{cgshu@shao.ac.cn}

\begin{abstract}

Quasar photometric redshifts are essential for studying cosmology and large-scale structures. However, their complex spectral energy distributions cause significant redshift–color degeneracy, limiting the accuracy of traditional methods. To overcome this, we introduce LSTM-MDNz, a novel end-to-end deep learning model combining long short-term memory networks (LSTM) with mixture density networks (MDN). The model directly uses multi-band photometric fluxes and associated errors as wavelength-ordered sequential inputs, eliminating the need for manual feature engineering while enabling simultaneous point estimation and probability distribution function (PDF) prediction of quasar redshifts. We integrate data from four major sky surveys—SDSS, DESI-LS, WISE, and GALEX—to assemble a sample of over 550,000 spectroscopically confirmed quasars ($0 \leq z_{\mathrm{spec}} \leq 5$) across 14 ultraviolet to infrared bands for model training and testing. Experimental results show that using all 14 bands yields optimal performance, with a normalized median absolute deviation ($\sigma_{\mathrm{NMAD}}$) of 0.037 and an outlier rate ($f_{\mathrm{out}}$) of 3.5\% on the test set. These values represent reductions of 29\% and 56\%, respectively, compared to the commonly adopted SDSS+WISE band set. Probability integral transform ($\mathrm{PIT}$) and continuous ranked probability score ($\mathrm{CRPS}$) analyses confirm that the predicted PDFs align closely with the true redshift distribution. Band-ablation experiments further highlight the essential role of ultraviolet and infrared data in alleviating color degeneracy and reducing systematic bias. This study demonstrates the effectiveness of multi-band fusion in improving quasar photo-z accuracy and offers a ready-to-use estimation framework for future surveys like LSST, CSST, and Euclid.

\end{abstract}



\section{Introduction} \label{sec:intro}

Quasars, the most energetically luminous active galactic nuclei in the universe, serve as vital probes for investigating the large-scale structure of the universe, galaxy evolution, and the growth of supermassive black holes \citep{1998ARA&A..36..267R,2010_HopkinsElvis_MNRAS.401....7H,2013ARA&A..51..511K}. They also play a crucial role in elucidating the nature of dark energy and understanding the epoch of cosmic reionization \citep{2006ARA&A..44..415F,2013PhR...530...87W,2016ApJ...828...26M}. The advent of next-generation large-scale photometric surveys, such as the Legacy Survey of Space and Time (LSST), the Chinese Space Station Survey Telescope (CSST), and the Euclid space telescope, is expected to discover and identify tens of millions of quasars \citep{2019ApJ...873..111I,2019ApJ...883..203G,2020A&A...642A.191E}. However, obtaining precise spectroscopic redshifts ($z_{\mathrm{spec}}$) for such a vast sample is observationally expensive and time-consuming, making spectroscopic confirmation impractical for the majority of sources. Consequently, efficient estimation of quasar photometric redshifts ($z_{\mathrm{phot}}$) using readily available multi-band photometric data has become an indispensable approach for statistical studies of large samples \citep{2001AJ....122.1151R,2004ApJS..155..243W}.

Photometric redshift estimation refers to the technique of inferring cosmological redshifts from multi-band photometric measurements (e.g., magnitudes and colors) without requiring spectroscopic observations. This method enables the rapid acquisition of three-dimensional spatial distributions for large samples, expanding the scale of observable universe studies by orders of magnitude compared to spectroscopic samples, thereby enhancing the statistical power of cosmological measurements \citep{hildebrandt2010phat}. Nearly all current and upcoming large-scale survey projects—including the Sloan Digital Sky Survey (SDSS; \citealt{1996AJ....111.1748F,2000AJ....120.1579Y}), the Dark Energy Survey (DES; \citealt{collaboration2016more,abbott2021dark}), the Kilo-Degree Survey (KiDS; \citealt{2013ExA....35...25D}), LSST \citep{abell2009lsst,2019ApJ...873..111I}, the Euclid Space Telescope \citep{laureijs2011euclid}, the Wide Field Infrared Survey Telescope or Nancy Grace Roman Space Telescope (WFIRST; \citealt{2012arXiv1208.4012G,spergel2015widefield,2019arXiv190205569A}), and the Hyper Suprime-Cam Subaru Strategic Program (HSC-SSP; \citealt{2018PASJ...70S...4A})—rely heavily on photometric redshifts to achieve their core scientific objectives.

Current photometric redshift estimation methods are broadly categorized into two types: template-fitting methods and machine learning methods. Template-fitting techniques such as HyperZ \citep{2011ascl.soft08010B}, BPZ \citep{2000ApJ...536..571B}, EAZY \citep{2008ApJ...686.1503B}, and Le Phare \citep{1999MNRAS.310..540A} derive redshifts by fitting photometric data with spectral energy distribution (SED) templates \citep{1996Natur.381..759L,1999ApJ...513...34F,2000A&A...363..476B,2024MNRAS.531.3539L}. Although this approach has a clear theoretical foundation, its accuracy heavily depends on the completeness and accuracy of the template library. In contrast, machine learning (or training-set) methods learn the complex mapping between photometric features and redshifts from large spectroscopic training samples, employing a wide variety of models. Commonly used techniques include artificial neural networks (ANN; \citealt{2003MNRAS.339.1195F,collister2004ANNz,2016PASP..128j4502S}), support vector machines (SVM; \citealt{2005PASP..117...79W}), self-organizing maps \citep{2012PASP..124..274W,2012MNRAS.419.2633G}, Gaussian process regression \citep{2006ApJ...647..102W}, k-nearest neighbors (kNN; \citealt{2007ApJ...663..774B}), boosted decision trees \citep{2010ApJ...715..823G}, random forests (RF; \citealt{2013MNRAS.432.1483C,2015MNRAS.452.3710R,2021MNRAS.502.2770M,2024MNRAS.52712140L}), and  sparse Gaussian frameworks \citep{2016MNRAS.455.2387A}. More recently, deep learning approaches such as multi-layer perceptrons (MLP; \citealt{2021ApJ...909...53Z}), convolutional neural networks (CNN; \citealt{2015MNRAS.452.4183H,2018A&A...609A.111D,2019A&A...621A..26P}), and Bayesian neural networks (BNN; \citealt{zhou2022photometricBNN}) have also been widely applied in photo‑$z$ estimation. When sufficient training samples are available, machine learning methods often achieve higher accuracy.

In recent years, with the growing volume of astronomical data and advances in computational capabilities, machine learning methods have been widely applied to photometric redshift estimation, delivering impressive results across various surveys. For instance, in galaxy photometric redshifts, \citet{2021MNRAS.502.2770M} used DES data and the COSMOS2015 catalog to show that random forest algorithms yield robust redshift estimates even with limited photometric bands. The Euclid team \citep{2020A&A...644A..31E} systematically evaluated 9 different machine learning methods on simulated galaxy data and found that most can provide reliable redshift estimates. To further improve performance, \citet{2024MNRAS.535.1844L} proposed an LSTM-based model for the CSST survey, which captures sequential correlations in multi-band data and uses Monte Carlo dropout to generate probability distribution functions (PDFs), thereby enhancing reliability. Additionally, studies by \citet{zhou2022photometricBNN} and \citet{2022MNRAS.512.1696H} demonstrated that multimodal approaches combining photometric and image data (e.g., hybrid MLP-CNN models) can further refine redshift and PDF estimation accuracy.

However, research on quasar photometric redshift estimation remains relatively limited. Compared to galaxies, quasar photometric redshifts generally exhibit lower accuracy and reliability, primarily due to their distinctive physical characteristics: quasar SEDs arise from a combination of power-law continuum emission, broad-line region clouds, dusty tori, and host galaxy light, leading to complex spectral shapes. Moreover, quasars lack pronounced features such as the Balmer break, the 4000Å break, or the Lyman break that characterize galaxy SEDs. These factors result in a highly nonlinear and non-monotonic relationship between quasar colors and redshifts—a phenomenon known as ``redshift–color degeneracy'' \citep{2001AJ....122.1151R,2022ARA&A..60..363N}. Consequently, methods that perform well for galaxy redshifts often show significantly reduced effectiveness when applied directly to quasars, facing limitations in both accuracy and robustness.

To improve quasar redshift estimation, researchers have explored diverse machine learning strategies. For example, \citet{2022MNRAS.509.2289L} developed models using XGBoost, CatBoost, and Random Forest, finding that a two-step strategy—separating quasars into high- and low-redshift subsets before prediction—outperformed single-model approaches. Their subsequent studies \citep{2023MNRAS.518..513L,2024AJ....168..233L} further indicated that these methods generally achieve higher accuracy for galaxies than for quasars. \citet{2021MNRAS.503.2639C} evaluated deep learning (DL), k-nearest neighbors (kNN), and decision tree regression (DTR) for quasar redshift estimation, concluding that kNN and DL delivered superior performance. \citet{2018A&A...611A..97P} converted quasar light curves into images and applied a convolutional neural network (CNN) to estimate redshifts for SDSS Stripe 82 quasars. Meanwhile, \citet{2023MNRAS.523.5799Y} proposed Q-PreNet, a network that integrates image and photometric data for quasar redshift estimation, showing promising results in single-value redshift prediction.

In the photometric redshift estimation, besides single-value (point) estimates, probability distribution functions (PDFs) also serve as an important form of output. Research shows that relying solely on point estimates and their error ranges often fails to fully capture the uncertainties in redshift estimation, while the complete probability distribution can provide a more reliable foundation for subsequent scientific analysis. For instance, in cosmological parameter estimation, using the full probability distribution of redshifts instead of a single value has been proven to effectively improve measurement accuracy \citep{2008JCAP...08..006M, 2009MNRAS.399.2279M}. Therefore, PDFs are particularly important in precision cosmological studies such as weak gravitational lensing and galaxy clustering \citep{2016PhRvD..94d2005B}.

Nevertheless, studies on PDF estimation for quasar photometric redshifts are still relatively scarce. Most current machine learning methods provide only single-point estimates and struggle to produce reliable PDFs, limiting their utility in precision cosmology. This gap is now being addressed by the mixture density network (MDN; \citealt{bishop1994mixture}). By combining the nonlinear fitting capability of neural networks with the probabilistic framework of Gaussian mixture models (GMM), MDN can directly learn and output complex conditional probability distributions, providing both accurate point estimates and associated uncertainties. For instance, \citet{2018A&A...609A.111D} integrated MDN with a deep convolutional network to propose the DCMDN architecture, which effectively estimates redshift PDFs from quasar images. \citet{2024AJ....168..244Z} further enhanced this approach through cross-modal fusion of photometric attributes (e.g., magnitudes and colors) with image features, achieving excellent PDF estimation performance using a CNN-based residual network combined with MDN.

Although integrating image data through CNNs and MDNs can enhance the performance of photometric redshift estimation, this approach demands substantially more computational resources than methods utilizing only photometric attributes (e.g., flux, magnitude, and color). Data preparation also presents challenges, as the collection and preprocessing of images require significant time and storage space. Furthermore, astronomical images often contain background noise and contaminants, which can adversely affect model performance. Consequently, image-based techniques are generally only feasible for smaller datasets and are difficult to scale to the vast data volumes anticipated from modern sky surveys.

In this study, we propose a novel end-to-end deep learning model, LSTM-MDNz, designed to deliver high-precision point estimates and PDFs for quasar photometric redshifts. The model combines long short-term memory (LSTM) networks with a mixture density network (MDN) and uses multi-band photometric fluxes (and their errors) as input, eliminating the need for manual feature engineering such as color indices. By arranging photometric fluxes and errors in wavelength order as sequential inputs, the LSTM module automatically captures inter-band dependencies and spectral feature correlations. This approach avoids the subjectivity and redundancy of manual feature selection, significantly streamlining input processing.

This study systematically trained and validated the LSTM-MDNz model using a sample of over 550,000 spectroscopically confirmed quasars. By integrating photometric data from four major surveys—the Sloan Digital Sky Survey (SDSS), the DESI Legacy Imaging Surveys (DESI-LS), the Wide-field Infrared Survey Explorer (WISE), and the Galaxy Evolution Explorer (GALEX)—we constructed a comprehensive multi-band dataset covering 14 bands from ultraviolet to infrared. Innovatively, photometric errors were incorporated as input features during modeling to enhance estimation performance. To further evaluate the contribution of different bands to redshift estimation, we conducted in-depth band-ablation experiments to quantitatively analyze the impact of various band combinations on measurement accuracy. The experimental results demonstrate the critical role of ultraviolet and infrared data in mitigating color degeneracy and reducing systematic biases. Furthermore, while maintaining high redshift estimation accuracy, the LSTM-MDNz model exhibited exceptional computational efficiency, characterized by low resource consumption and fast training speed, indicating its strong applicability and scalability for large-scale astronomical data processing.

The remainder of this paper is organized as follows: Section \ref{sec:dataset} describes the multi-band photometric data and sample construction; Section \ref{sec:method} details the network architecture and training strategy of the LSTM-MDNz model; Section \ref{sec:performance} presents experimental results and analysis, including model performance evaluation, band-ablation studies, and key findings; and Section \ref{sec:summary} summarizes the research outcomes and provides discussion.

\section{Data} \label{sec:dataset}

The dataset utilized in this study is derived from the 16th Data Release Quasar Catalog of the Sloan Digital Sky Survey (SDSS DR16Q; \citealt{2020_Lyke_ApJS..250....8L}). This catalog encompasses 750,414 spectroscopically confirmed quasars, each accompanied by celestial coordinates (Right Ascension and Declination), spectroscopic identifiers (MJD, PLATE, FIBERID), spectroscopic redshifts (\texttt{spec-z}), and multi-band photometric data from cross-match with other survey programs.

For each quasar in the SDSS DR16Q catalog, we retrieved its photometric data across the five optical bands ($u$, $g$, $r$, $i$, $z$) from the SDSS DR16 photometric catalog through coordinate-based cross-matching. To enhance optical wavelength coverage, this study also incorporated photometric measurements in the $g$, $r$, and $z$ bands from the DESI Legacy Imaging Surveys Data Release 9 (DESI-LS DR9; \citealt{2021AAS...23723503S}), based on the official online cross-matching catalog between SDSS DR16Q and DESI-LS DR9 \footnote{\url{https://portal.nersc.gov/cfs/cosmo/data/legacysurvey/dr9/north/external/}} \footnote{\url{https://portal.nersc.gov/cfs/cosmo/data/legacysurvey/dr9/south/external/}}, which adopts a matching radius of 1.5 arcseconds. While these bands partially overlap with those from SDSS, the differences in filter systems provide complementary color information for our machine learning model. This also allows us to evaluate its generalization capability when SDSS $u$- and $i$-band data are not available.

The official cross‑match catalog between SDSS DR16Q and DESI‑LS DR9 is divided into the Northern and Southern galactic caps, with some overlap between the two regions: the Northern cap includes 361,449 quasars, and the Southern cap 430,727. Given the consistency in observational strategy, depth, and data quality between the two caps in DESI-LS, we merged the two datasets to simplify processing. Objects appearing in both caps were retained as independent observational entries to maximize data usage. Tests confirmed that key performance metrics derived from the merged sample were consistent with those from the individual cap samples, indicating that the merging process did not significantly affect model performance.

DESI DR9 also provides forced-photometry infrared fluxes in the $W1$–$W4$ bands, measured on coadded unWISE images \citep{2019ApJS..240...30S} using the TRACTOR package \citep{2016ascl.soft04008L}. This forced-photometry approach recovers infrared fluxes for faint sources by leveraging the precise optical positions from DESI, thereby substantially improving sample completeness compared to standard WISE catalog-based methods (e.g., ALLWISE or CatWISE).

The final training sample includes model fluxes and associated errors in 14 bands: the five SDSS optical bands ($u$, $g$, $r$, $i$, $z$); the three DESI-LS optical bands ($g$, $r$, $z$); the two ultraviolet bands ($FUV$, $NUV$) from GALEX, as included in SDSS DR16Q; and the four WISE infrared bands. This multi-wavelength dataset, covering ultraviolet to infrared, supports a comprehensive analysis of quasar spectral energy distributions.

\begin{figure} 
        \includegraphics[width=0.6\textwidth]{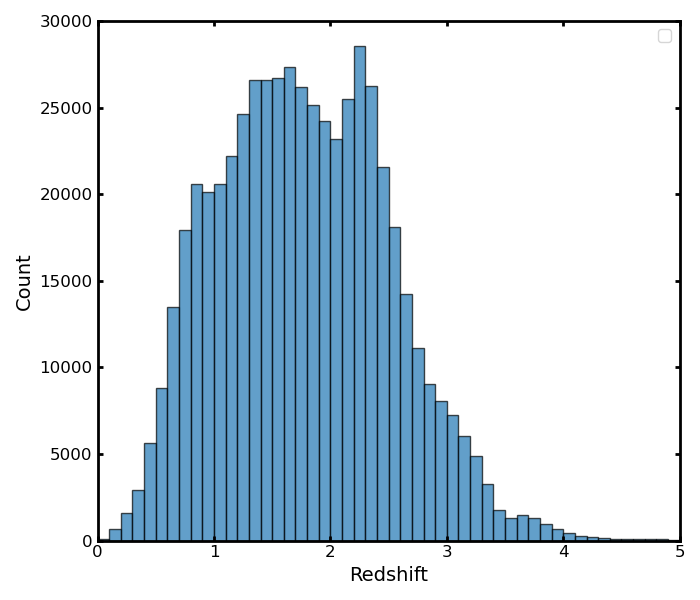} 
        \centering
        \caption{Redshift distribution of the quasar sample with complete 14-band photometry. The vertical axis shows the number of sources, and the horizontal axis the redshift.} 
        \label{fig:redshift} 
\end{figure}

Due to the small number of very high-redshift ($z > 5$) quasars—insufficient for robust statistical training—we restricted the sample to $z \leq 5$, following the approach of \citet{2018A&A...609A.111D} and \citet{2023MNRAS.523.5799Y}. We also excluded sources with unreliable spectroscopic redshifts (i.e., $z_{\mathrm{warning}} \neq 0$). After filtering, the final sample consisted of 558,111 quasars with complete photometry across all 14 bands. Figure~\ref{fig:redshift} shows the redshift distribution of the sample.

\section{Methodology} \label{sec:method}

In this section, we introduce LSTM-MDNz, a novel end-to-end deep learning model specifically designed for estimating quasar photometric redshifts. The model integrates LSTM networks with MDNs, requiring only multi-band photometric fluxes and their corresponding errors as input. It simultaneously provides high-precision photometric redshift point estimates and complete PDFs. We begin by separately introducing the basic architecture of LSTM recurrent neural networks and MDNs, followed by a detailed description of the overall network design and training methodology for LSTM-MDNz.

\subsection{Long Short-Term Memory (LSTM) Network}

The LSTM network is a specialized type of recurrent neural network (RNN) first introduced by \citet{1997_Hochreiter_LSTM}. It has since been extensively extended and optimized, becoming a mainstream model for processing sequential data. By incorporating a memory cell state, the LSTM effectively mitigates the vanishing and exploding gradient problems that traditional RNNs often encounter with long sequences. Owing to its strong memory capacity and ability to capture long-range dependencies, LSTM has been widely applied in areas such as time series forecasting, reliability prediction, and natural language processing \citep{2014arXiv1409.3215S,2022ApJ...930...70H,2023ApJ...954..164T}. In recent years, LSTM and other RNN architectures have also gained traction in astronomical research, demonstrating particular promise in photometric redshift estimation \citep{2024MNRAS.535.1844L,2024A&C....4900886T}.

The fundamental unit of an LSTM shares structural similarities with a traditional RNN but employs a more sophisticated gating mechanism. Each LSTM unit contains a memory block, which maintains and updates information across time steps via three gates: the forget gate, the input gate, and the output gate, along with a cell state. These components work together to regulate information flow in a fine-grained manner.

Specifically, at time step $t$, the LSTM unit takes the current input $x_t$  and the previous hidden state $h_{t-1}$, and computes the following:

\begin{equation}
f_t = \sigma\left( W_f \cdot [h_{t-1}, x_t] + b_f \right),
\label{eq:forget_gate}
\end{equation}

\begin{equation}
i_t = \sigma\left( W_i \cdot [h_{t-1}, x_t] + b_i \right),
\label{eq:input_gate}
\end{equation}

\begin{equation}
o_t = \sigma\left( W_o \cdot [h_{t-1}, x_t] + b_o \right),
\label{eq:output_gate}
\end{equation}

\begin{equation}
\tilde{C}t = \tanh \left( W_C \cdot [h{t-1}, x_t] + b_C \right),
\label{eq:candidate_state}
\end{equation}
where $f_t$, $i_t$ and $o_t$ represent the forget, input, and output gates, respectively; $\tilde{C}_t$ is the candidate cell state; $\sigma$ is the sigmoid function; and $\mathrm{tanh}$ is the hyperbolic tangent function. The weight matrices
$W_f$, $W_i$, $W_o$, $W_C$ and bias terms $b_f$, $b_i$, $b_o$, $b_C$ are shared across time steps.

The cell state $C_t$ and hidden state $h_t$ are then updated as:
\begin{equation}
C_t = f_t \odot C_{t-1} + i_t \odot \tilde{C}_t,
\label{eq:cell_update}
\end{equation}

\begin{equation}
h_t = o_t \odot \tanh(C_t),
\label{eq:hidden_update}
\end{equation}
where $\odot$ denotes element-wise multiplication. Initial states are typically set as $C_0 = 0$ and $h_0 = 0$.

In this mechanism, the forget gate controls how much of the previous cell state is retained; the input gate regulates the integration of new candidate information; and the output gate governs the exposure of the cell state to the hidden output. This gating system allows the LSTM to maintain long-term dependencies and process long sequences effectively, mitigating gradient issues common in vanilla (basic) RNNs.

Compared to conventional machine learning methods, LSTM can automatically learn complex temporal dependencies without manual feature engineering. They excel at capturing long-range patterns and dynamic evolution in sequential data. For example, in natural language processing, LSTMs effectively model long-distance semantic relationships, which improves performance in tasks such as machine translation \citep{1997_Hochreiter_LSTM,2013arXiv1303.5778G}. Similarly, in photometric redshift estimation, using LSTM to model multi-band photometric sequences allows the model to capture nonlinear relationships across bands and spectral variations with wavelength, thereby improving prediction accuracy and robustness.

In this study, we employ an LSTM to model multi-band photometric sequences for redshift estimation. This approach enables the model to capture nonlinear relationships between different bands and the variation of flux with wavelength, which is expected to improve the accuracy and stability of predictions. In contrast, a MLP treats the input as an unordered set of features and struggles to utilize the natural order of wavelengths. This limitation hinders its ability to identify local features (e.g., absorption lines) and global trends (e.g., continuum slopes) \citep{2024MNRAS.535.1844L}.

{To further enhace context capture, we consider a} bidirectional LSTM (Bi-LSTM), an extension of the standard LSTM. A Bi-LSTM processes the input sequence in both forward and reverse directions using two separate hidden layers. This architecture allows the model to incorporate both past and future context at every time step, capturing bidirectional dependencies. Bi-LSTMs often improving performance in sequence modeling tasks such as time series forecasting and language understanding \citep{GRAVES2005602,2015arXiv150801991H} and have shown promise in astronomical applications \citep{2024MNRAS.535.1844L}. Leveraging these advantages, the LSTM-MDNz model proposed in this paper adopts a Bi-LSTM architecture.

\subsection{Mixture Density Network (MDN)}

The mixture density network (MDN), introduced by \citet{bishop1994mixture}, is a neural network architecture designed for regression tasks where the relationship between inputs and outputs may be multimodal or inherently uncertain. Unlike conventional regression models that output a single deterministic value, an MDN models the full conditional probability distribution $p(y|x)$, allowing it to capture complex uncertainty patterns and multiple plausible outcomes for a given input.

An MDN combines the expressive power of a neural network with the flexibility of a mixture model—typically a Gaussian mixture model (GMM). Instead of directly predicting the target variable, the network outputs the parameters of a mixture distribution. For a univariate output $y$ and $K$ Gaussian components, the MDN predicts three sets of parameters per input $x$: the mixture weights $w_k$, the means $\mu_k$, and the variances $\sigma_k^2$ of each Gaussian component. The mixture weights must be non-negative and sum to 1, the means determine the central position of each component, and the variances control the dispersion of the distributions. 

Thus, the output layer must produce $3K$ values. The resulting conditional density is given by:
\begin{equation}
p(y \mid x) = \sum_{k=1}^{K} w_k(x) \mathcal{N}\left( y \mid \mu_k(x), \sigma_k^2(x) \right),
\label{eq:mdn_p}
\end{equation}
where $\mathcal{N}$ denotes the Gaussian probability density function. All parameters - the weights $w_k$, means $\mu_k$, and variances $\sigma_k^2$ - are functions of the input $x$, learned by the neural network.

Architecturally, an MDN generally consists of a feature-extracting backbone (e.g., a multilayer perceptron, MLP) followed by a final linear layer that outputs the $3K$ distribution parameters. To enforce constraints: the mixture weights $w_k$ are passed through a softmax activation to ensure they are positive and sum to 1, the means $\mu_k$ are typically unbounded and emitted directly, and the variances $\sigma_k^2$ must be positive; this is usually achieved by applying an exponential activation to the corresponding network outputs.

Training is performed via maximum likelihood estimation. For a ground-truth pair$(x,y)$, the loss is the negative log-likelihood:
\begin{equation}
\mathcal{L} = -\log p(y \mid x).
\label{eq:mdn_loss}
\end{equation}

Minimizing this loss over the training set encourages the MDN to match the predicted distribution to the true data distribution.

Due to its ability to output complete probability distributions, the MDN excels in tasks that require the quantification of predictive uncertainty. In recent years, it has garnered significant attention in the field of astronomical photometric redshift estimation, with relevant studies demonstrating its potential in handling data uncertainty and providing reliable redshift estimates \citep{2018A&A...609A.111D,2024AJ....168..244Z, 2024A&C....4900886T}.

\subsection{LSTM-Augmented Mixture Density Network (LSTM-MDNz)}

To effectively estimate quasar redshifts from multi-band photometric data and provide reliable uncertainty quantification, this study draws inspiration from the hybrid architecture proposed by \citet{2018A&A...609A.111D} and develops a deep fusion model that integrates LSTM with MDN, referred to as LSTM-MDNz. In contrast to previous approaches, our model does not utilize quasar image data but instead relies solely on multi-band photometric sequences for redshift inference. This model leverages the strength of LSTMs in handling long-term dependencies and dynamic feature extraction in sequential data while incorporating the capability of MDNs for modeling complex conditional probability distributions. This enables the model to directly output comprehensive PDFs of redshift values based on multi-band photometric data, rather than providing a single point estimate. This design effectively addresses uncertainties and multi-modal issues in redshift estimation, significantly enhancing the interpretability and practical value of the results.

\begin{figure} 
        \includegraphics[width=\textwidth]{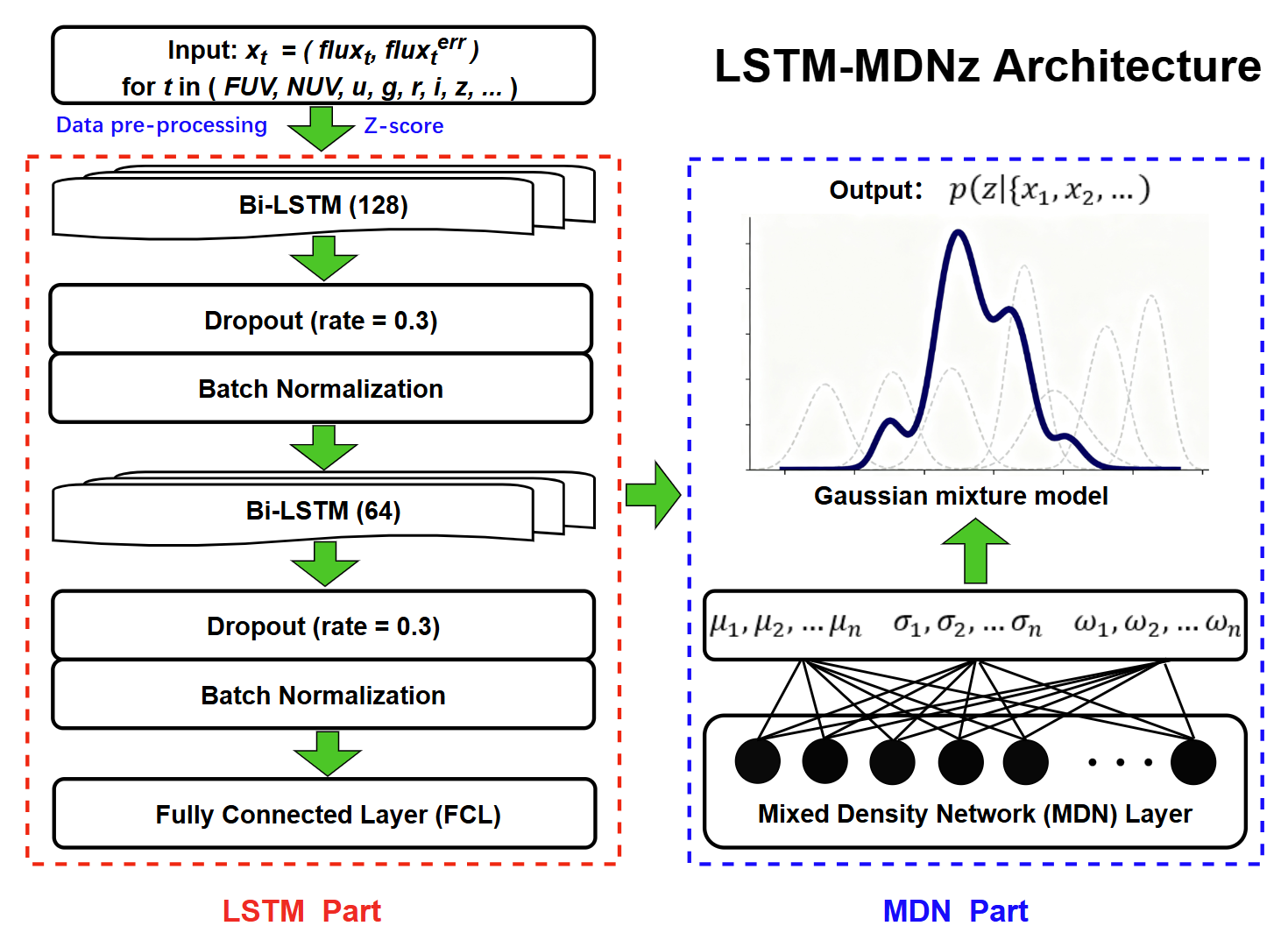} 
        \caption{Architecture and data flow of the LSTM-MDNz model. The input consists of multi-band photometric fluxes and their corresponding errors. These sequences are processed by the LSTM encoding module (red dashed box), which includes bidirectional LSTM layers, dropout, and batch normalization for feature extraction and stabilization. The resulting features are transformed via a fully connected layer before being passed to the MDN module (blue dashed box), where a Gaussian mixture model is used to compute the conditional probability density function (PDF) of the photometric redshift.} 
        \label{fig:framework} 
\end{figure}

As illustrated in Figure \ref{fig:framework}, the LSTM-MDNz model consists of two core components: a deep bidirectional LSTM feature extractor and an MDN output layer. The feature extractor includes two stacked bidirectional LSTM layers that process the multi-band photometric sequences in a hierarchical manner. The first Bi-LSTM layer captures local dependencies and basic temporal patterns, while the second layer learns more abstract representations and long-range contextual relationships. This deep bidirectional encoding structure enables the model to integrate both forward and backward information, enhancing its ability to perceive spectral energy distribution (SED) shapes and their variations, thereby comprehensively modeling complex dependencies in the input photometric sequences.

The final hidden states from the last Bi-LSTM layer are aggregated into a unified feature vector, which is then passed through a fully connected layer for nonlinear transformation and further abstraction. This high-level feature vector serves as input to the MDN layer. Based on a Gaussian mixture model, the MDN outputs three sets of parameters for each component $k$: the mixture weight $w_k$, mean $\mu_k$, and standard deviation $\sigma_k$. These parameters are used to reconstruct the conditional probability distribution $p(z|x)$ of the redshift $z$, enabling probabilistic inference and uncertainty-aware redshift estimation.

\subsection{Training the LSTM-MDNz} \label{subsec:training}

During model training, the photometric data for each quasar are first transformed into a vector sequence. Specifically, for each band, the measured flux  $Flux_t$  and its corresponding error $Flux_t^{err}$ are combined into a two-dimensional vector $(Flux_t, Flux^{err}_t)$. These vectors are then arranged in ascending order of wavelength to form the input sequence for the LSTM-MDNz model.

To improve training stability, each feature dimension (i.e., flux or error values for a particular band across all training samples) is standardized as follows:
\begin{equation}
x^* = \frac{x - \mu}{\sigma},
\label{eq:input_standardized}
\end{equation}
where $x$ is the original value, and $\mu$ and $\sigma$ are the mean and standard deviation of that feature computed over the training set. This standardization ensures that all input features are on a comparable scale, which aids gradient-based optimization and model convergence.

The standardized sequence is then passed to the LSTM encoding module (see Figure \ref{fig:framework}), which consists of two bidirectional LSTM layers. The first Bi-LSTM layer contains 128 hidden units per direction, and the second contains 64 units per direction. Each Bi-LSTM layer is followed by a dropout layer with a rate of 0.3 and a batch normalization layer. Dropout serves as a regularization technique to reduce overfitting by randomly disabling neuron connections during training \citep{2012arXiv1207.0580H,2014arXiv1409.2329Z}. Batch normalization stabilizes and accelerates training by normalizing layer inputs within each mini-batch \citep{2015arXiv150203167I}. The output of the final batch normalization layer is aggregated into a feature vector and passed through a fully connected layer before entering the MDN module.

The MDN module models the conditional probability distribution $P(z|x)$ of the redshift $z$ given the photometric sequence $x={x_1,x_2,... x_T}$ using a Gaussian mixture model. {The number of Gaussian components is set to 15, which was determined through systematic validation experiments comparing different counts (ranging from 5 to 25). This value optimally balances model flexibility and generalization performance, yielding the lowest negative log-likelihood on the validation set without overfitting.} The network is trained by minimizing the negative log-likelihood loss (Equation~\ref{eq:mdn_loss}), which jointly optimizes the LSTM feature extractor and the MDN parameters to accurately fit the redshift distribution.

The model is implemented using the TensorFlow and Keras frameworks. Training is performed on an NVIDIA L40S GPU with a batch size of 128. We use the Adam optimizer \citep{kingma2014adam} with hyperparameters $\beta_1 = 0.9$, $\beta_2 = 0.999$, and an initial learning rate of 0.001. The learning rate is reduced by a factor of 10 every 50 epochs. Early stopping is applied if the validation loss does not improve for 20 consecutive epochs, with the best model weights being restored. The full training process using all bands takes approximately 75 minutes to complete. To ensure full reproducibility and facilitate community use, the source code, training scripts, and associated data are publicly available in our GitHub repository\footnote{\url{https://github.com/zjluo-code/LSTM-MDNz}}.

\section{Experiments and Results} \label{sec:performance}

This section details the training and evaluation of the LSTM-MDNz model using the quasar sample described in Section \ref{sec:dataset}, with the aim of assessing its accuracy and reliability in photometric redshift estimation. The full sample is partitioned into training, validation, and test sets using a 7:1.5:1.5 ratio. Model training uses only the training set, while the validation set is used for real-time performance monitoring during training. Early stopping is triggered if the validation loss shows no improvement over 20 consecutive epochs to prevent overfitting. All final evaluations are performed exclusively on the test set.

\subsection{Evaluation Metrics}

The primary goal of our model is to produce high-precision point estimates of quasar photometric redshifts along with accurate PDFs. Accordingly, the evaluation metrics are divided into two categories: point estimate statistics, which assess the accuracy of the predicted spectroscopic redshift values, and PDF evaluation metrics, which gauge the reliability and statistical quality of the predicted probability density functions.

For point estimate statistics, we employ several widely adopted metrics: mean squared error (MSE), mean absolute error (MAE), catastrophic outlier fraction ($f_{\mathrm{out}}$), normalized median absolute deviation ($\sigma_{\mathrm{NMAD}}$), and bias ($\mathrm{bias}$). MSE and MAE are standard regression metrics that quantify the average discrepancy between the true and estimated values:
\begin{equation}
\mathrm{MSE} = \frac{1}{N} \sum_{i=1}^{N} (z_{\mathrm{spec}}^i - z_{\mathrm{phot}}^i)^2,
\label{eq:z_mse}
\end{equation}
\begin{equation}
\mathrm{MAE} = \frac{1}{N} \sum_{i=1}^{N} |z_{\mathrm{spec}}^i - z_{\mathrm{phot}}^i|,
\label{eq:z_mae}
\end{equation}
where $N$ is the total number of samples, $z_{\rm spec}^i$ is the spectroscopic redshift (true value) of the $i$-th sample, and $z_{\rm phot}^i$ is the corresponding photometric redshift estimate.

The catastrophic outlier fraction $f_{\mathrm{out}}$ measures the proportion of severely incorrect estimates. Following \citet{2018A&A...619A..14F} and \citet{2020A&A...644A..31E}, an estimate is classified as a catastrophic outlier if:
\begin{equation}
\frac{|z_{\mathrm{spec}} - z_{\mathrm{phot}}|}{1 + z_{\mathrm{spec}}} > 0.15.
\label{eq:outlier_cond}
\end{equation}
Such outliers indicate that the corresponding photometric redshift estimate deviates significantly from the true value.

The normalized median absolute deviation ($\sigma_{\mathrm{NMAD}}$) evaluates the overall accuracy of the estimates and is defined as \citep{2008ApJ...686.1503B}:
\begin{equation}
\sigma_{\mathrm{NMAD}} = 1.48 \times \mathrm{median} \left( \frac{|\Delta z|}{1 + z_{\mathrm{spec}}} \right),
\label{eq:nmad_def}
\end{equation}
 where $\Delta z =z_{\rm phot}-z_{\rm spec}$. $\sigma_{\mathrm{NMAD}}$ is preferred over standard deviation due to its robustness to outliers. The factor 1.48 ensures that under a normal distribution, $\sigma_{\mathrm{NMAD}}$ corresponds to the standard deviation.

Bias quantifies the systematic deviation in the estimates, indicating an overall tendency to over- or underestimate the true redshifts:
\begin{equation}
\mathrm{bias} = \mathrm{median} \left( \frac{z_{\mathrm{phot}} - z_{\mathrm{spec}}}{1 + z_{\mathrm{spec}}} \right).
\label{eq:bias_def}
\end{equation}
This metric helps identify systematic overestimation or underestimation.

For PDF assessment, we use the continuous ranked probability score (CRPS) and the probability integral transform (PIT) to evaluate the consistency between the predicted and true distributions \citep{2018A&A...609A.111D}. CRPS is a proper scoring rule that measures the accuracy of probabilistic forecasts by comparing the predicted cumulative distribution function (CDF) against the observed value. For a predictive CDF $F$ and an observation $x$, CRPS is defined as:
\begin{equation}
\mathrm{CRPS}(F, x) = \int_{-\infty}^{+\infty} \left[ F(y) - \mathbf{1}_{\{y \geq x\}} \right]^2 dy,
\label{eq:crps_def}
\end{equation}
where $y$ represents a variable in the integral, and $\mathbf{1}_{\{y \geq x\}}$ is the Heaviside indicator function.

PIT assesses the calibration quality of probabilistic forecasts by evaluating how well the predicted distribution represents the true uncertainty \citep{dawid1984present,2018A&A...609A.111D,2020A&A...644A..31E,2021MNRAS.502.2770M}. For each object, the PIT value is the predicted CDF evaluated at the true redshift:
\begin{equation}
\mathrm{PIT} = \int_{-\infty}^{z_{\mathrm{spec}}} \mathrm{PDF}(z) dz,
\label{eq:pit_def}
\end{equation}
where $z_{\rm{spec}}$ is the true redshift, and $\mathrm{PDF}(z)$ is the probability density function at redshift $z$.

For well-calibrated PDFs, the PIT values across the sample should follow a standard uniform distribution $U(0,1)$ \citep{dawid1984present}. This arises because if the PDF is correctly calibrated, the true redshift can be viewed as a random draw from it. Deviations from uniformity in the PIT distribution indicate systematic biases: a U-shaped distribution suggests under-dispersed PDFs, while an inverted U-shape implies over-dispersion. Slope variations in the PIT distribution may also reveal systematic errors \citep{2016arXiv160808016P}. PIT has been widely used in redshift PDF validation studies (e.g., \citealt{2010MNRAS.406..881B,10.1093/pasj/psx077,2021MNRAS.502.2770M}).

\subsection{Model Performance with All Available Bands}

This subsection presents the performance of the LSTM-MDNz model when all 14 photometric bands are available. This configuration serves as the benchmark scenario in our study, aimed at evaluating the model’s optimal capability given complete multi-band photometric information. The results will provide a baseline for subsequent band-ablation experiments.

The input features—the photometric flux and its corresponding error for each band—are combined into two-dimensional vectors, which are then arranged in ascending order of wavelength to form a sequence of length 14. In the case of overlapping coverage between the SDSS and DESI-LS bands, the SDSS data is placed before the DESI-LS data in the sequence. Experiments indicate that the specific ordering of bands has negligible effects on the model output. The model architecture follows the description in Section \ref{subsec:training}, utilizing a two-layer bidirectional LSTM for feature extraction and employing a MDN with 15 Gaussian components to output the conditional redshift distribution. The final photometric redshift point estimate ($z_{\rm phot}$) is taken as the mean of the predicted PDF.

Table~\ref{tab:metrics} summarizes the point estimation performance on the training, validation, and test sets. Under full-band conditions, the model achieves excellent accuracy on the test set: the normalized median absolute deviation ($\sigma_{\mathrm{NMAD}}$) is 0.037, the catastrophic outlier fraction ($f_{\rm out}$) is 3.4\%, and the $\rm bias$ is nearly zero (-0.0006). The consistency of metrics across all datasets indicates strong generalization without overfitting.

\begin{table}[ht]
\centering
\caption{Performance metrics of the LSTM-MDNz model for photometric redshift point estimation using all 14 photometric bands.}
\label{tab:metrics}
\setlength{\tabcolsep}{8pt}
\renewcommand{\arraystretch}{1.2}
\begin{tabular}{lcccccc}
\hline
        \textbf{Data set} & \textbf{No. of objects} & $\bm{\sigma_{\rm NMAD}}$ & $\bm{f_{\rm out} (\%)}$ & $\bm{bias}$ & \textbf{MSE} & \textbf{MAE} \\
        \hline
        Training Set & 390,677 & 0.036 & 3.2 & -0.0004 &  0.034 & 0.105 \\
        Validation set & 83,717 & 0.038 & 3.5 & -0.0006 & 0.040 & 0.112 \\
        Test set & 83,717 & 0.037 & 3.4 & -0.0006 & 0.039 & 0.110 \\
\hline
\end{tabular}
\end{table}

\begin{figure} 
    \centering
    \includegraphics[width=0.8\textwidth]{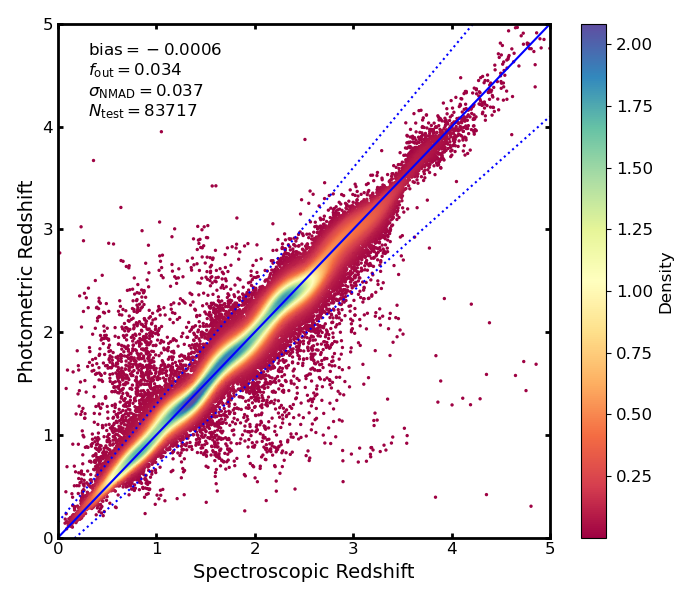} 
    \caption{Photometric vs. spectroscopic redshifts for test set samples under full-band (14 bands) input. The solid line is the diagonal; dashed lines show $|z_{\mathrm{spec}} - z_{\mathrm{phot}}|/(1 + z_{\mathrm{spec}})=0.15$. Color indicates point density.} 
    \label{fig:zsp_zph} 
\end{figure}

Figure~\ref{fig:zsp_zph} illustrates a scatter density distribution, comparing photometric redshifts with spectroscopic redshifts for samples in the test set. This comparison provides a visual assessment of the accuracy and reliability of photometric redshift estimates. As shown in the figure, most data points align closely along the diagonal, indicating strong agreement between predicted (photometric) and true (spectroscopic) redshift values. Importantly, this high degree of consistency is maintained across the full redshift range ($0 \leq z \leq 5$), with no discernible systematic deviations. This consistent performance demonstrates the robustness and accuracy of our photometric redshift estimation method.

\begin{figure} 
    \centering
    \includegraphics[width=0.8\textwidth]{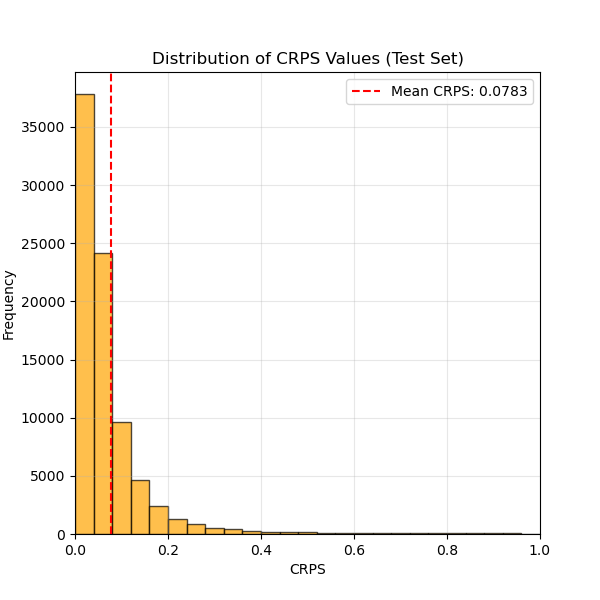} 
    \caption{ Distribution of CRPS values for the test set under full-band (14 bands) input. The red dashed line marks the mean CRPS.} 
    \label{fig:crps} 
\end{figure}

The distribution of the CRPS across the entire test set is shown in Figure~\ref{fig:crps}, with a mean value of approximately 0.0783. CRPS is calculated by integrating the squared difference between the CDF of the predicted probabilities and the Heaviside indicator function of the true redshift value, as defined in Equation~\ref{eq:crps_def}. This relatively low mean value indicates that the predicted PDFs closely align with the observed spectroscopic redshift values, thereby validating the model’s capability in accurately assessing the sharpness of the PDFs.

\begin{figure} 
    \centering
    \includegraphics[width=\textwidth]{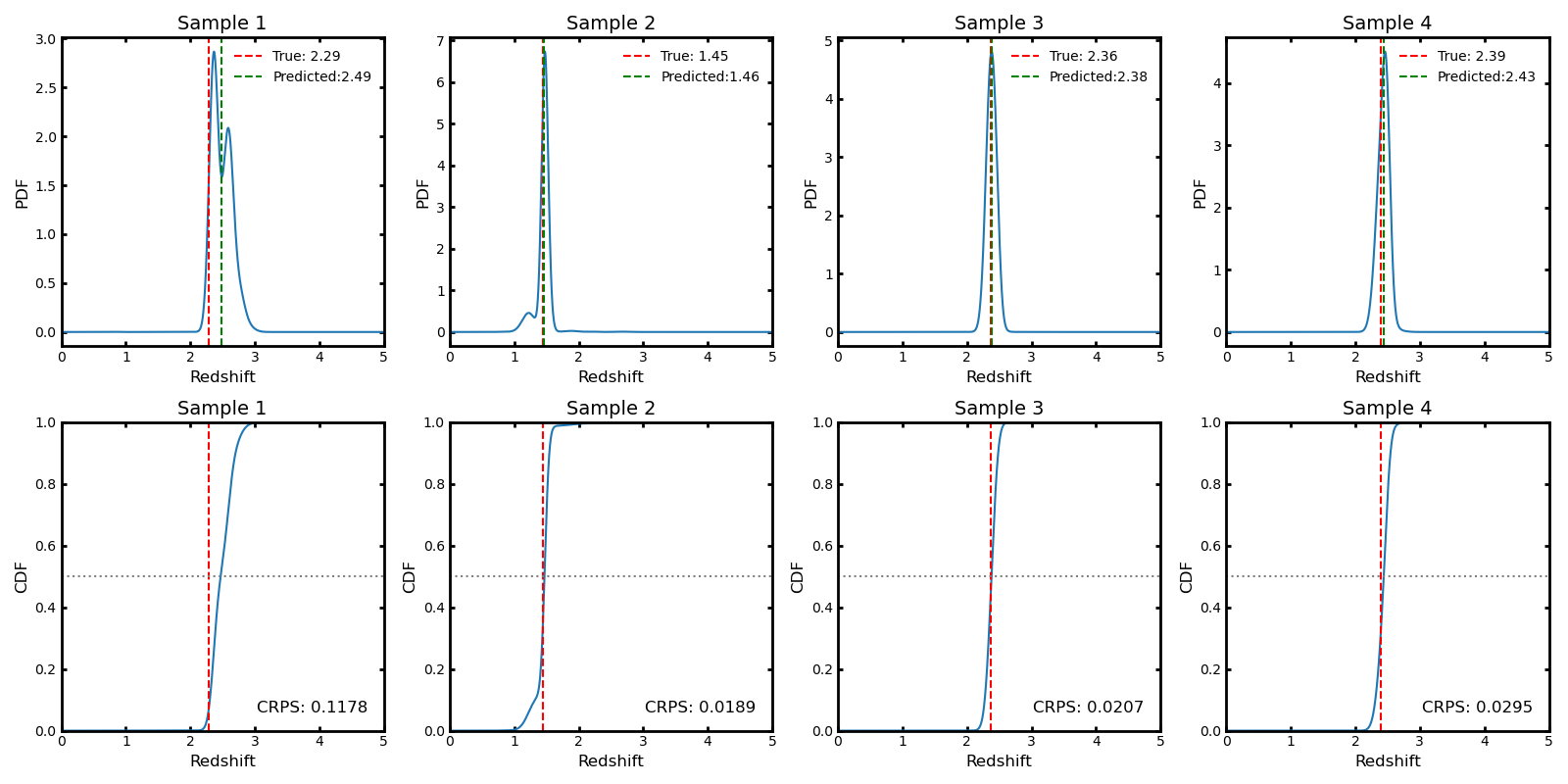} 
    \caption{Examples of photometric redshift predictions for four randomly selected quasars under full-band (14 bands) input. The solid blue line in the upper panel represents the predicted PDF, while the solid blue line in the lower panel shows the corresponding CDF. The red dashed line indicates the position of the true spectroscopic redshift ($z_{\mathrm{spec}}$), and the blue dashed line represents the photometric redshift ($z_{\mathrm{phot}}$).} 
    \label{fig:pdf_cdf} 
\end{figure}

As depicted in Figure~\ref{fig:pdf_cdf}, we present PDFs (upper panel) and their corresponding CDFs (lower panel) for four randomly selected quasars, drawn from our test set. Each subplot is marked with a red dashed vertical line indicating the true spectroscopic redshift ($z_{\rm spec}$) of the respective quasar. A detailed examination of the plots reveals that most observed spectroscopic redshift values lie within the high-probability regions of their predicted PDFs. This outcome signifies not only the model’s proficiency in generating precise point predictions but also its exceptional ability to produce reliable uncertainty quantifications. The consistent proximity between predicted and true redshifts corroborates the robustness and accuracy of our methodology.

\begin{figure} 
    \centering
    \includegraphics[width=\textwidth]{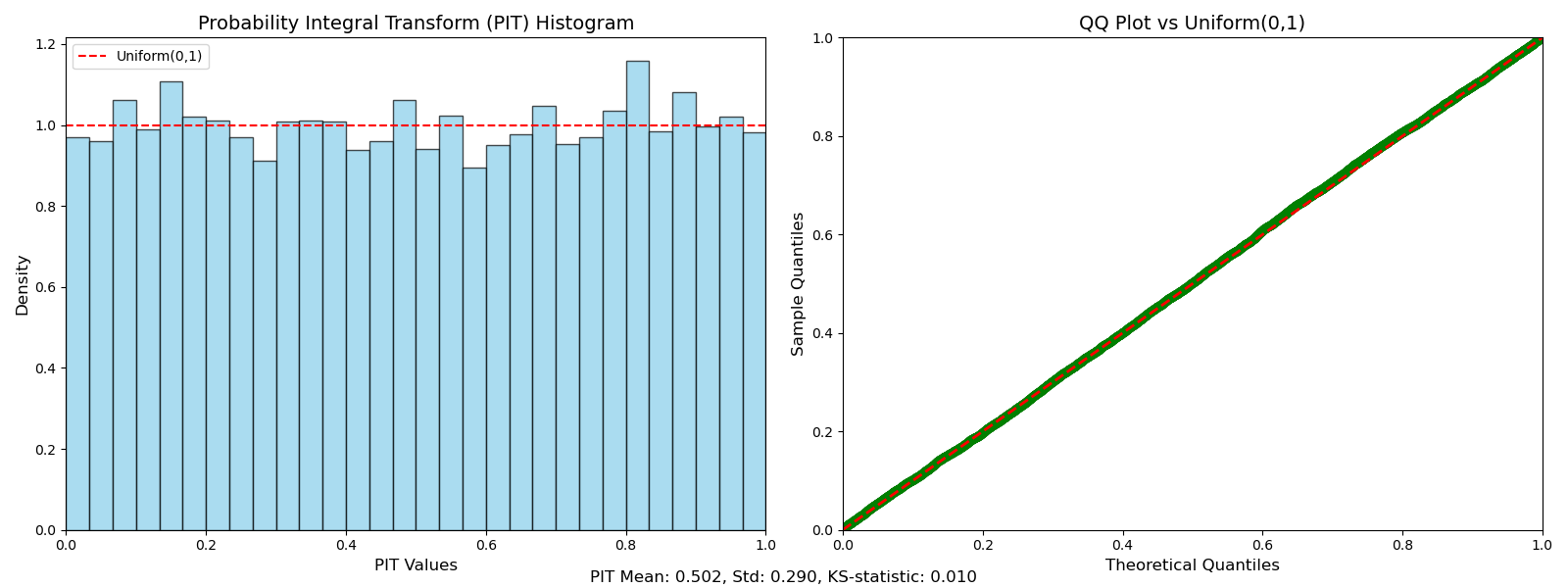} 
    \caption{PIT histogram (left) and Q–Q plot (right) for the test set under full-band (14 bands) input. In the left panel, the blue histogram represents the observed PIT distribution, and the red dashed line indicates the reference line for the ideal uniform distribution (U(0,1)). In the right panel, the scatter points show the correspondence between the observed quantiles and theoretical quantiles, with the red dashed line representing the theoretical $y=x$ line.}
    \label{fig:pit} 
\end{figure}

The left panel of Figure~\ref{fig:pit} illustrates the PIT distribution for our test set samples. A perfectly calibrated model would produce PIT values uniformly distributed over the interval $[0,1]$, as indicated by the red horizontal dashed line in the figure. The actual PIT distribution of our samples, shown here as a blue histogram, closely resembles this uniform distribution, with only minor localized deviations.

The right panel of Figure~\ref{fig:pit} presents a quantile-quantile (Q-Q) plot comparing the observed PIT values to those expected under a uniform $(0,1)$ distribution. In an ideally calibrated model, all points would lie along the red reference line representing the theoretical quantiles. As demonstrated in this figure, our model exhibits excellent calibration, with most points cluster tightly around the reference line, indicating that the statistical properties of the PIT values are highly consistent with a uniform distribution.

This strong agreement between the observed and expected distributions validates not only the accuracy of our point estimates but also the reliability of the uncertainty quantification provided by the model. The close adherence to uniformity in both the histogram and Q-Q plot underscores the robustness of our approach in capturing the inherent uncertainties in redshift estimation.

Our evaluation demonstrates that fusing 14 photometric bands, ranging from ultraviolet to infrared, effectively samples the spectral energy distribution (SED). This enables the model to better capture the characteristics of the SED shape of quasars and to break the redshift-color degeneracy.  These advances yield not only more accurate point estimates but also better-calibrated probabilistic predictions. The resulting uncertainties faithfully represent the true error distribution, which is crucial for scientific applications like cosmological parameter inference.

\subsection{Band-ablation Experiments}

To systematically evaluate the contributions of various band data to the accuracy of quasar photometric redshift estimation and to confirm the critical role of ultraviolet (GALEX) and infrared (WISE) bands in resolving the redshift-color degeneracy, we conducted a series of systematic band-ablation experiments. These experiments were designed to assess how the absence of specific bands affects model performance by examining different combinations of input data while keeping model architecture, hyperparameters, and training strategies consistent.

The experiments employed six distinct band configurations:

1. All Bands (14 bands): Serving as the baseline, this configuration incorporates data from all available surveys: GALEX ultraviolet ($FUV$, $NUV$), SDSS optical ($u$, $g$, $r$, $i$, $z$), DESI-LS optical ($g$, $r$, $z$), and WISE infrared ($W1$, $W2$, $W3$, $W4$).

2. Optical Only (8 bands): This setup uses only optical bands from SDSS and DESI-LS, simulating scenarios where ultraviolet and infrared data are unavailable.

3. Optical + WISE (12 bands): Combining optical bands with WISE infrared data while excluding GALEX ultraviolet observations.

4. Optical + GALEX (10 bands): Merging optical bands with GALEX ultraviolet data while omitting WISE infrared contributions.

5. SDSS + WISE (9 bands): Reproducing a combination frequently used in previous studies \citep{2023MNRAS.523.5799Y,2024AJ....168..244Z}, comprising SDSS optical bands and WISE infrared data.

6. DESI + WISE (7 bands): Representing scenarios where specific SDSS filters (particularly $u$ and $i$ bands) are unavailable, testing the capability of DESI Legacy Survey optical data combined with WISE infrared observations.

All experiments employed identical training, validation, and test datasets to ensure result comparability. For point estimation accuracy, we focused on changes in two key metrics: normalized median absolute deviation ($\sigma_{\rm NMAD}$) and the catastrophic outlier rate ($f_{\rm out}$). Regarding probabilistic predictions, our analysis concentrated primarily on the mean CRPS to assess the sharpness and reliability of the predicted PDFs.

Table \ref{tab:band_com} summarizes the performance of the LSTM-MDNz model on the test set under different band combinations. The experimental results clearly demonstrate the importance of various band data for accurate photometric redshift estimation. When utilizing all 14 bands, the model achieves superior performance, with $\sigma_{\rm NMAD} = 0.037$, $f_{\rm out} = 3.4\%$, and mean $\mathrm{CRPS} = 0.0783$. By comparison, when only the 8 optical bands from SDSS+DESI are used, model performance significantly deteriorates: $\sigma_{\rm NMAD}$ increases to 0.068, $f_{\rm out}$ rises to 20.6\%, and the mean CRPS worsens to 0.1657. This result aligns with the redshift-color degeneracy challenge discussed in the introduction, highlighting that optical information alone is insufficient for uniquely determining quasar redshifts.

When infrared or ultraviolet data were added to the optical bands (SDSS + DESI), the model’s performance showed significant improvement in both scenarios. Specifically, after incorporating WISE infrared data, the normalized median absolute deviation ($\sigma_{\rm NMAD}$) decreased to 0.041, representing an approximate 40\% improvement compared to the configuration using only optical bands. Additionally, the catastrophic outlier rate ($f_{\rm out}$) was dramatically reduced to 5.4\%, indicating an approximate 74\% improvement, while the CRPS decreased to 0.0911, reflecting around a 45\% enhancement.

In the case of adding GALEX ultraviolet data, $\sigma_{\rm NMAD}$ dropped to 0.047, which corresponds to approximately 31\% improvement compared to the optical-only configuration. The outlier rate significantly decreased to 9.2\%, equating to around a 55\% improvement, and the mean CRPS reduced to 0.1123, indicating approximately a 32\% enhancement. These findings demonstrate that both infrared and ultraviolet data provide crucial information not captured by optical bands alone, enabling the model to better distinguish between quasars with similar optical colors but differing redshifts due to degeneracy.

\begin{table}[ht]
    \centering
    \caption{Performance of the LSTM-MDNz Model on Photometric Redshift Estimation Across Different Band Combinations}
    \label{tab:band_com}
    \setlength{\tabcolsep}{8pt} 
    \renewcommand{\arraystretch}{1.2} 
    \begin{tabular}{lccccccc}
        \hline
        \textbf{Band combination} & \textbf{No. of bands} & $\bm{\sigma_{\rm NMAD}}$ & $\bm{f_{\rm out} (\%)}$ & $\bm{bias}$ & \textbf{MSE} & \textbf{MAE} & \textbf{CPRS (mean)} \\
        \hline
        All Bands & 14 & 0.037 & 3.4 & -0.0006 &  0.039 & 0.110 & 0.0783 \\
        Optical + WISE & 12 & 0.041 & 5.4 & -0.0005 & 0.056 & 0.129 & 0.0911\\
        Optical + GALEX & 10 & 0.047 & 9.2 & -0.0010 & 0.086 & 0.163 & 0.1123 \\
        SDSS + WISE & 9 & 0.052 & 7.9 & -0.0010 & 0.073 & 0.159 & 0.1106\\
         DESI + WISE & 7 &0.080 & 16.8 & -0.0017 & 0.137 & 0.235 & 0.1635\\
        Optical Only & 8 & 0.068 & 20.6 & 0.0050 & 0.169 & 0.258 & 0.1657 \\
        \hline
    \end{tabular}
    \vspace{0.3cm}
\end{table}

Furthermore, a comparison of model performance across different combinations (Table \ref{tab:band_com}) indicates that optimal results are achieved only when all bands (ultraviolet, optical, and infrared) are employed together. The performance advantage of the “All Bands” configuration over the “Optical + WISE” or “Optical + GALEX” combinations demonstrates that the complementary information provided by ultraviolet and infrared bands is not entirely redundant; rather, it exhibits a synergistic effect that collectively enhances the model’s accuracy.

%Furthermore, a comparison of the performance across different combinations reveals that optimal results are achieved only when all bands (ultraviolet, optical, and infrared) are used together. The performance advantage of the "All Bands" configuration over either the "Optical + WISE" or "Optical + GALEX" combinations indicates that the complementary information provided by ultraviolet and infrared bands is not entirely redundant but exhibits a synergistic effect, jointly enhancing the model's accuracy.

Table \ref{tab:band_com} also illustrates an interesting phenomenon: although both SDSS and DESI-LS belong to the optical band category, their differing filter response functions allow for the integration of complementary optical data from different survey projects, which can significantly enhance model performance. For instance, compared to the SDSS+WISE combination, the Optical + WISE configuration (i.e., SDSS + DESI + WISE) achieves a remarkable 21\% reduction in the normalized median absolute deviation ($\sigma_{\rm NMAD}$) from 0.052 to 0.041, a 34\% decrease in the outlier rate ($f_{\rm out}$) from 7.9\% to 5.4\%, and an 18\% reduction in the mean CRPS from 0.1106 to 0.0911. Conversely, when the optical band coverage is reduced—such as by replacing SDSS+WISE with DESI + WISE, which results in the loss of $u$-band and $i$-band information—the model’s performance in photometric redshift prediction declines noticeably.

For the SDSS + WISE combination commonly adopted in previous studies, the performance achieved by the LSTM-MDNz model in this paper is $\sigma_{\rm NMAD} = 0.052$, $f_{\rm out} = 7.9\%$, and mean $\mathrm{CRPS} = 0.1106$. This result significantly surpasses the performance reported by \citet{2021MNRAS.503.2639C}, who utilized similar SDSS + WISE photometric data ($\sigma_{\rm NMAD} = 0.103$, $f_{\rm out} = 24\%$). The improvement can likely be attributed to our probabilistic LSTM-MDN architecture, which directly models the sequential nature of multi-band data and explicitly quantifies redshift uncertainties through mixture density modelling. It also outperforms the quasar redshift estimation results obtained by \citet{2023MNRAS.523.5799Y} using their Q-PreNet model that integrates image and photometric features ($\sigma_{\rm NMAD} = 0.070$, $f_{\rm out} = 13.7\%$, mean $\mathrm{CRPS}$ = 0.1318). Notably, this advancement is achieved despite our model using only photometric sequences, highlighting the effectiveness of our LSTM-based encoding. Furthermore, our model’s performance is comparable to the PDF estimation results achieved by \citet{2024AJ....168..244Z} through cross-modal fusion of photometric and image features using CNNs and MDNs, with a mean CRPS of 0.1187.

However, when compared to the 14-band fusion scheme proposed in this paper, labeled as All Bands, the SDSS + WISE combination exhibits a significant performance gap. Specifically, the $\sigma_{\rm NMAD}$ value increases by approximately 41\%, the catastrophic outlier rate $f_{\rm out}$ rises by about 132\%, and the mean CRPS increases by approximately 41\%. This comparison underscores the critical importance of further integrating DESI-LS, which provides complementary optical information, and GALEX, offering ultraviolet data, on top of the SDSS and WISE data to achieve current high-precision photometric redshifts.

\begin{figure}
    \centering
    \begin{subfigure}[b]{0.44\textwidth}
        \centering
        \includegraphics[width=\textwidth]{zsp_zph_all.png}
        \caption{All Bands}
        \label{fig:zz1}
    \end{subfigure}
    \hspace{0.0\textwidth} 
    %\hfill % Add some space between the subfigures
    \begin{subfigure}[b]{0.44\textwidth}
        \centering
        \includegraphics[width=\textwidth]{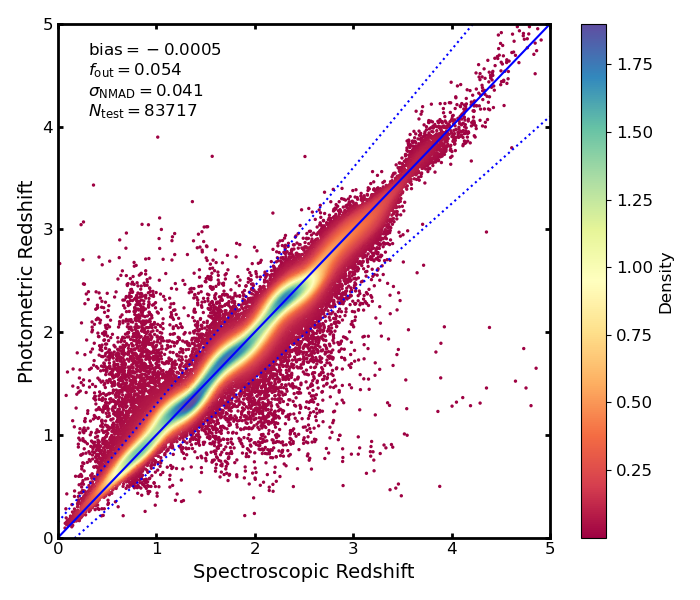}
        \caption{Optical + WISE}
        \label{fig:zz2}
    \end{subfigure}
    \vskip 0.0\baselineskip
    %\vskip\baselineskip % Add some vertical space between rows

    \begin{subfigure}[b]{0.44\textwidth}
        \centering
        \includegraphics[width=\textwidth]{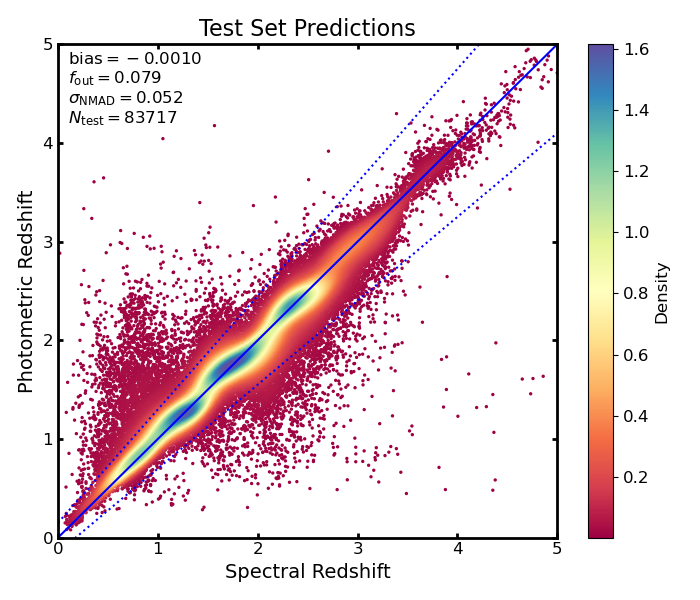}
        \caption{SDSS + WISE}
        \label{fig:zz3}
    \end{subfigure}
    \hspace{0.0\textwidth}
    %\hfill % Add some space between the subfigures
    \begin{subfigure}[b]{0.44\textwidth}
        \centering
        \includegraphics[width=\textwidth]{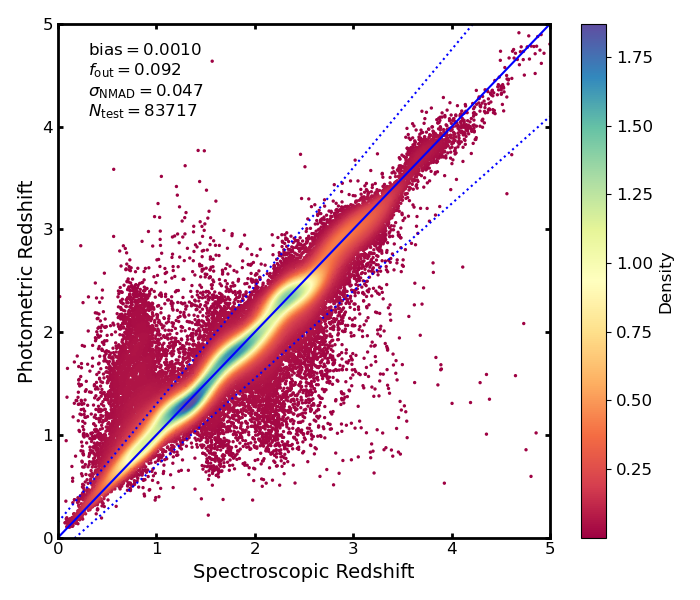}
        \caption{Optical + GALEX}
        \label{fig:zz4}
    \end{subfigure}
    \vskip 0.0\baselineskip
    %\vskip\baselineskip % Add some vertical space between rows

    \begin{subfigure}[b]{0.44\textwidth}
        \centering
        \includegraphics[width=\textwidth]{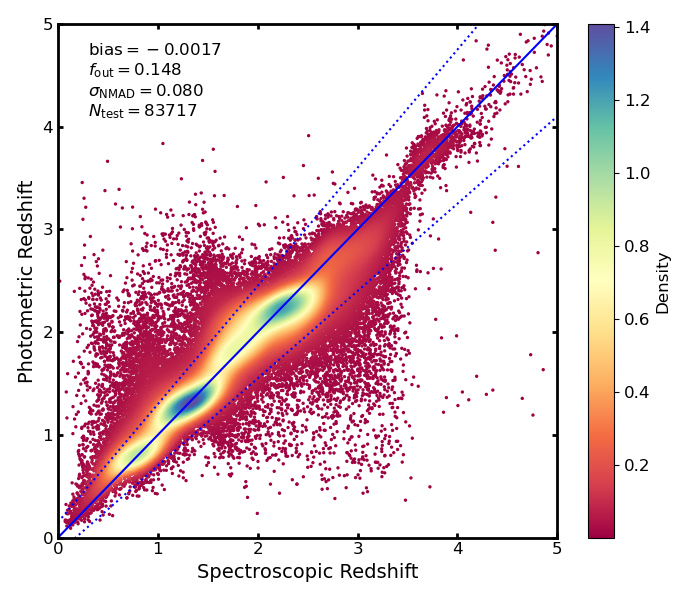}
        \caption{DESI + WISE}
        \label{fig:zz5}
    \end{subfigure}
    \hspace{0.0\textwidth}
    %\hfill % Add some space between the subfigures
    \begin{subfigure}[b]{0.44\textwidth}
        \centering
        \includegraphics[width=\textwidth]{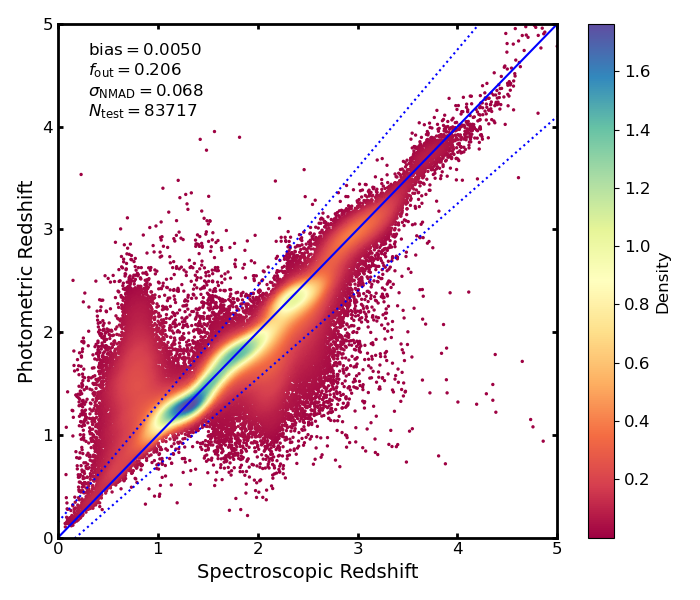}
        \caption{Optical Only}
        \label{fig:zz6}
    \end{subfigure}
    \caption{Spectroscopic redshifts versus predicted photometric redshifts derived from the proposed LSTM-MDNz model for the test set under different band combinations.}
    \label{fig:combined}
\end{figure}

Figure~\ref{fig:combined} illustrates the scatter density distributions of photometric versus spectroscopic redshifts under various band combinations. Observations reveal that as certain bands are excluded, particularly ultraviolet or infrared data, vertical stripe-like patterns in the scatter plots become more pronounced. These stripes indicate increased redshift confusion due to color degeneracy. Additionally, both systematic bias and dispersion of photometric redshifts significantly increase when critical bands are absent, highlighting the importance of comprehensive band coverage for accurate redshift estimation.

Through our comprehensive band-ablation experiments, we have demonstrated that infrared (WISE) and ultraviolet (GALEX) bands play pivotal roles in addressing the redshift–color degeneracy of quasars. These bands not only mitigate systematic biases but also significantly enhance the accuracy and reliability of photometric redshift estimations. In addition, the integration of complementary optical information from diverse survey projects, such as the DESI Legacy Surveys (DESI-LS), provides substantial support for improved model performance.

\section{Summary and Discussion} \label{sec:summary}

%This study addresses the challenges of accuracy and reliability in quasar photometric redshift estimation caused by redshift-color degeneracy by proposing a novel end-to-end deep learning model, LSTM-MDNz. The model innovatively integrates long short-term memory networks with mixture density networks, requiring only multiband photometric fluxes and their measurement errors as input, without the need for manual feature engineering, to simultaneously obtain high-precision point estimates of redshifts and their complete probability distribution functions. By organizing photometric data from different bands into sequential inputs according to wavelength order, the model fully leverages the strengths of LSTM in sequence modeling, enabling it to automatically capture interdependencies between bands and structural features of spectral energy distributions. On this basis, the MDN further maps the learned high-dimensional features into parameters of a Gaussian mixture distribution, thereby achieving probabilistic modeling of redshift uncertainty.

This study addresses the fundamental challenge of redshift-color degeneracy in quasar photometric redshift estimation by proposing LSTM-MDNz, an end-to-end framework that integrates long short-term memory networks with mixture density networks. The model processes multiband photometric fluxes and their measurement errors to simultaneously generate both precise point estimates and complete probability distribution functions, eliminating manual feature engineering while providing comprehensive uncertainty quantification.

The model's architecture is specifically designed to capture spectral dependencies by structuring photometric data as wavelength-ordered sequences. The LSTM component extracts spectral energy distribution features and inter-band relationships, while the MDN layer maps these features to parameters of a Gaussian mixture distribution. This integrated approach enables robust probabilistic modeling of redshift uncertainties, effectively addressing limitations inherent in traditional estimation methods.

The LSTM-MDNz model demonstrates good performance on a quasar dataset covering 14 bands from ultraviolet to infrared. In terms of point estimation, it achieves a normalized median absolute deviation ($\sigma_{\rm NMAD}$) of 0.037, a catastrophic outlier rate ($f_{\rm out}$) as low as 3.4\%, and a bias close to zero. For probabilistic prediction, statistical analysis of the PIT indicates that the predicted photometric redshift PDFs are well-calibrated, while the CRPS score reveals a high degree of correspondence between the predicted PDFs and the spectroscopically measured redshift values, demonstrating accurate and reliable PDF uncertainty estimation. These results validate the strong capability of the LSTM architecture in processing multi-band photometric sequences and capturing spectral energy distribution features, as well as the effectiveness of the MDN in modeling complex conditional probability distributions.

Through systematic band-ablation experiments, we have quantitatively demonstrated the crucial role of ultraviolet (GALEX) and infrared (WISE) bands in addressing the redshift-color degeneracy issue in quasars. Compared to using optical bands alone, incorporating either ultraviolet or infrared data significantly enhances model performance—reducing $\sigma_{\rm NMAD}$ by approximately 30-40\%, lowering $f_{\rm out}$ (catastrophic outlier rate) by 55-75\%, and decreasing CPRS by 30-45\%. Additionally, integrating complementary optical band data from different survey projects also yields significant performance gains for the model. The ultimate optimal performance arises from a synergistic effect across all bands (ultraviolet, optical, and infrared) rather than simple additive effects. This underscores the importance of constructing broadly covered multi-band datasets for robust and accurate modeling.

%Through systematic band ablation experiments, we have quantitatively demonstrated the indispensable role of ultraviolet (GALEX) and infrared (WISE) bands in resolving the redshift-color degeneracy problem in quasars. Compared to using optical bands alone, the incorporation of either ultraviolet or infrared data significantly improves model performance (approximately 30\%-40\% improvement in $\sigma_{\rm NMAD}$ and 50\%-70\% reduction in $f_{\rm out}$). Furthermore, integrating complementary optical band data from different survey projects also brought substantial performance gains to the model. The ultimate optimal performance results from the synergistic effect of all bands (ultraviolet, optical, and infrared) rather than simple additive effects, highlighting the importance of constructing broadly covered multi-band datasets.

Compared with existing studies, the LSTM-MDNz model in this research has notable advantages. When using similar SDSS and WISE benchmark photometric data, it can match and even surpass the performance of complex models that require multimodal data, which validates the model's good ability to extract key information from photometric data. The core value of the model lies in its end-to-end modeling approach. It relies solely on photometric data and does not require manual feature engineering, thus having great potential for large-scale applications. We believe that LSTM-MDNz provides a powerful and scalable solution to deal with the vast amount of data from next-generation sky surveys such as DESI, LSST, Euclid, and CSST.

%Compared to existing studies, the LSTM-MDNz model proposed in this work demonstrates certain advantages. When evaluated on the same SDSS+WISE dataset, using only photometric data, it achieves performance comparable to and even partially superior to some complex models that incorporate image features, indicating that the model maintains relatively high accuracy while also exhibiting good computational efficiency. The characteristics of LSTM-MDNz—such as its end-to-end modeling capability, elimination of manual feature engineering, and high computational efficiency—suggest its potential for processing the massive datasets generated by next-generation large-scale sky surveys. Therefore, this study provides a referential framework for quasar redshift estimation in major sky survey projects such as DESI, LSST, Euclid, and CSST.

While the LSTM-MDNz model demonstrates promising performance in quasar photometric redshift estimation, its capabilities and broader applicability can be further enhanced in several directions. Firstly, the current model is primarily validated on bright SDSS quasars; its performance on fainter sources with larger photometric errors—common in surveys like LSST and Euclid—requires further testing, as increased noise may exacerbate redshift-color degeneracy and affect PDF calibration. Future work will explore noise-aware training or transfer learning on deeper, noisier datasets to improve robustness. Secondly, generalization to higher redshifts ($z > 5$) is currently limited by scarce training data, but targeted optimization can be pursued as high-redshift samples expand. Thirdly, integrating additional data modalities (e.g., variability features or morphological parameters) could further boost accuracy and robustness. Lastly, systematic evaluation under varied observational depths and selection functions will be crucial for optimizing the model’s practical utility across diverse surveys.

%Although the LSTM-MDNz model demonstrates promising performance in quasar photometric redshift estimation, several aspects warrant further improvement. First, the current model primarily applies to quasars with redshifts $z \leq 5$, while its generalization capability may be limited for high-redshift quasars ($z > 5$) due to scarce training samples. As more high-redshift quasar samples become available in the future, specialized optimization or training for this population could enhance the model's estimation capability at the high-redshift end. Second, while this study primarily relies on photometric data, future work could explore effectively integrating the LSTM-MDNz architecture with other modalities of quasar information (such as variability features and morphological parameters) to further improve the accuracy and robustness of redshift estimation. Additionally, applying the model to more real survey data and systematically evaluating its generalization ability under different observational depths and selection functions will help optimize its practicality and adaptability.

%% Please use the acknowledgment and contribution environments. This will 
%% be anonomyized when the "anonymous" style option is used. 
\begin{acknowledgments}
Z.L. acknowledges the support from the National Natural Science Foundation of China (Grant No. 12573009) and the scientific research grants from the China Manned Space Project with Grand No. CMS-CSST-2025-A07. L.F. acknowledges the support from the Innovation Program of Shanghai Municipal Education Commission (Grant No. 2025GDZKZD04) and China Manned Space Project with Grant No. CMS-CSST-2025-A05. S.Z. acknowledges support from the National Natural Science Foundation of China (Grant No. 12173026), the National Key Research and Development Program of China (Grant No. 2022YFC2807303), the Shanghai Science and Technology Fund (Grant No. 23010503900), the Program for Professor of Special Appointment (Eastern Scholar) at Shanghai Institutions of Higher Learning, and the Shuguang Program (23SG39) of the Shanghai Education Development Foundation and Shanghai Municipal Education Commission. H.X. acknowledges the support from the National Natural Science Foundation of China (NSFC 12203034), the Shanghai Science and Technology Fund (22YF1431500), and the Shanghai Municipal Education Commission regrading artifical intelligence empowered research. This work is also supported by the National Natural Science Foundation of China under Grant No. 12141302. 

\end{acknowledgments}

\bibliography{ref}{}
\bibliographystyle{aasjournalv7}



\end{document}